\documentstyle [12pt]{article}
\title{{\bf A study on the rare radiative decay \\   
$B_c \rightarrow D_s^*\gamma$ in technicolor with scalars}}
\author{ Lu Gongru, Cao Yigang, Xiong Zhaohua and Xiao Zhenjun  \\
\small Physics Department of Henan Normal University, Xinxiang, Henan, 
453002, P. R. China \\ }
\date{}

\begin{document}
\maketitle
\begin{abstract}	
Applying the perturbative QCD ( PQCD ), we study the process $B_c\rightarrow 
D_s^*\gamma$ in the technicolor with a massless 
scalar doublet model ( TCMLSM ) recently presented by C. D. Carone and H. 
Georgi; and compare the results with that estimated in the standard 
model ( SM ). 
There are two mechanisms which contribute to the $B_c\rightarrow D_s^*
\gamma$ process. One proceeds through the short distance $b\rightarrow s 
\gamma$ transition, the other through weak annihilation accompanied by a 
photon emission. In the SM, these two processes are found to contribute with 
the same order of magnitude. In the TCMLSM, the modification of 
$B_c\rightarrow D_s^*\gamma$ from $\pi_p$ ( the physical pions in this model 
) is small for the allowed mass of $\pi_p$.
\end{abstract}

\vspace{0.5cm}


\newpage
\begin{flushleft}    
\section*{I. Introduction}
\end{flushleft}

The inclusive rare decay $B\rightarrow X_s\gamma$ has been studied several 
years before [1]. Recently the physics of $B_c$ meson has caught intensive 
attentions [2]. The $B_c$ meson is believed to be the next and the final 
family of $B$ mesons. It provides unique opportunity to examine various 
heavy quark fragmentation models, heavy quark spin-flavor symmetry, different 
quarkonium bound state models and properties of inclusive decay channels. 
Being made of two heavy quarks of different flavors, $B_c$ radiative weak 
decay also offer a rich source to measure elements of Cabibbo-Kobayashi-
Maskawa ( CKM ) matrix of the standard 
model ( SM ). 

Different from the general rare $B$ decays $B\rightarrow X_s\gamma$ which is 
mainly induced by the flavor-changing $b\rightarrow s\gamma$ neutral currents 
[3], in $B_c\rightarrow D_s^*\gamma$, the bound state effects could 
seriously modify the results from the assumption. Bound state effects include 
modifications from weak annihilation which involve no neutral flavor-changing 
currents at all. The effects of weak annihilation mechanism are expected 
large due to the large CKM amplitude.  

Unfortunately, the well known chiral-symmetry [4] and the heavy quark 
symmetry [5] can not be applied to $B_c\rightarrow D_s^*\gamma$ process. 
Recently, a perturbative 
QCD ( PQCD ) analysis of $B$ meson decays seems to give a good prediction 
[6]. As it is argued in Ref.[7], $B_c$ two body nonleptonic decay can be 
conveniently studied within the framework of PQCD suggested by Brodsky-Lepage 
[8] and then developed in Ref.[6]. Here, we summarize their idea: 
In the subprocess $b\rightarrow s\gamma$, 
s quark obtains a large momentum by recoiling, in order to form a bound state 
with the spectator $\overline c$ quark, the most momentum of s must be 
transferred to $\overline c$ by a hard scattering process. In the final 
bound state ( i.e. $D_s^*$ ), since the heavy charm should share the most 
momentum of $D_s^*$, the hard scattering is suitable for PQCD calculation 
[6, 8].    

In $B_c\rightarrow D_s^* \gamma$, the 
subprocess $b\rightarrow s\gamma$,  taken as a free decay, is usually 
controlled by the one-loop electromagnetic penguin diagrams which are in 
particular sensitive to contributions from those new physics beyond the SM. 
This situation is similar with $B\rightarrow K^*\gamma$.

Recently, the modification of the electromagnetic penguin interaction 
to the decay $b\rightarrow s\gamma$ from PGBs in the one generation 
technicolor model 
( OGTM ) has been estimated in Ref.[9]. In the recent literatures [10, 11], 
the technicolor ( TC ) models with scalars have been 
studied extensively. The phenomenological studies have shown that the TC 
models with scalars do not generate unacceptably large FCNCs and are 
consistent with the 
experimental constraints on oblique electroweak radiative corrections. 
Among these models, the TC with a massless scalar doublet model ( TCMLSM ) 
[10], presented by C. D. Carone and H. Georgi
 is the simplest nontrivial extension of the SM with only two new free 
parameters
 ( $h_+$, $h_-$ or $h$, $\lambda$ ). The phenomenology of the TCMLSM has 
been discussed in the literatures [11]. In this paper we express these two 
new free parameters as $m_{\pi_p}$ and $\frac{f}{f'}$ differently from Ref.
[10]. The relation of them can be found in Ref.[10], where $m_{\pi_p}$ is the 
mass of $\pi_p$ ( the physical pions in this model ), $f$ and $f'$ are the 
technipion decay constant and scalar vacuum expectation value ( VEV ), 
respectively. The couplings of charged $\pi_p$ with ordinary fermions are 
given as  
$$         		
[\pi_p^+--u_i--d_j]=-\frac{\sqrt2}{2}V_{u_id_j}\frac{f}{vf'}[m_{d_j}(1
+\gamma_5)-m_{u_i}(1-\gamma_5)],
\eqno{(1)}
$$ 
$$                        
[\pi_p^---u_i--d_j]=-\frac{\sqrt2}{2}V_{u_id_j}\frac{f}{vf'}[m_{u_i}(1
+\gamma_5)-m_{d_j}(1-\gamma_5)],
\eqno{(2)}
$$
where u=( u, c, t ), d=( d, s, b ), 
$V_{u_id_j}$ is the 
element of CKM matrix, and $v\sim$ 250 GeV is the electroweak scale. 

In this paper, applying the above PQCD method, we address $B_c\rightarrow 
D_s^*\gamma$ in the TCMLSM to examine the virtual effects of 
$\pi_p$ in the TCMLSM and compare the results with which estimated in the 
SM. 

This paper is organized as follows: In Sec.II, we display our calculations 
in the SM and TCMLSM. We present the final numerical results in Sec.III. 
Sec.IV contains the discussion. 

\begin{flushleft}
\section*{II. Calculation}
\end{flushleft}

Using the factorization scheme [8] within PQCD, the momentum of quarks
are taken as some fractions $x$ of the total momentum of the meson weighted 
by a soft physics distribution functions $\Phi_{H}(x)$. The meson wave 
functions of $B_c$ and $D_s^*$ take the simple form of $\delta$ function ( 
the so-called peaking approximation ) [12, 13]:  
$$
\Phi_{B_c}(x)=\frac{f_{B_c}\delta(x-\epsilon_{B_c})}{2\sqrt{3}},
\eqno{(3)}
$$
$$
\Phi_{D_s^{*}}(y)=\frac{f_{D_s^{*}}\delta(y-\epsilon_{D_s^{*}})}{2\sqrt{3}}, 
\eqno{(4)}
$$
The normalization [13] is
$$
\int_{0}^{1} dx\Phi_{B_c}(x)=\frac{f_{B_c}}{2\sqrt{3}},
\eqno{(5)}
$$
$$
\int_{0}^{1} dy\Phi_{D_s^*}(y)=\frac{f_{D_s^*}}{2\sqrt{3}},
\eqno{(6)}
$$
where $x$, $y$ denote the momentum fractions of $c$, $s$ quarks in the $B_c$ 
and $D_s^*$ mesons, $f_{B_c}$ and $f_{D_s^*}$ are decay constants of $B_c$ 
and $D_s^*$ respectively,                
$$
\epsilon_{B_c}=\frac{m_c}{m_{B_c}},                  
\eqno{(7)}
$$
$$
\epsilon_{D_s^*}=\frac{m_{D_s^*}-m_c}{m_{D_s^*}}.
\eqno{(8)}
$$
The spinor part of $B_c$ and $D_s^{*}$ [14] are
$$
\frac{(\not p+m_{B_c})\gamma_5}{\sqrt{2}},
\eqno{(9)}
$$
$$  
\frac{(\not q-m_{D_s^*})\not \epsilon}{\sqrt{2}},         
\eqno{(10)}
$$
which come from the matrix structures of $B_c$ and $D_s^*$ meson wave 
functions, while $p$ and $q$ are the momenta of the $B_c$ and $D_s^*$ 
respectively, and $\epsilon$ is the polarization vector of $D_s^{*}$.

\begin{flushleft}
\subsection*{II ( i ). Electromagnetic penguin contribution}
\end{flushleft}

The relevant Feynman diagrams which contribute to the short distance 
electromagnetic penguin process $b\rightarrow s\gamma$ are illustrated as 
the blob of Fig.1. 

In the evaluation, we at first integrate out the top quark and the weak $W$ 
bosons at $\mu=m_W$ scale, generating an effective five-quark theory. By 
using the renormalization group equation, we run the effective field theory 
down to b-quark scale to give the leading log QCD corrections.

After applying the full QCD equations of motion [15], a complete set of 
dimension-6 operators relevant for $b\rightarrow s\gamma$ decay can be 
chosen as $O_1$ - $O_8$, which have been given in the Refs.[1, 9]. The 
effective Hamiltonian appears at the $W$ scale is given as  
$$
H_{eff}=\frac{4G_F}{\sqrt{2}}V_{tb}V_{ts}^* \sum\limits_{i=1}\limits^{8}C_i
(m_W)O_i(m_W).
\eqno{(11)}
$$
The coefficients of 8 operators are:
$$
C_i(m_W) = 0, i= 1, 3, 4, 5, 6, C_2(m_W) = -1,
\eqno{(12)}
$$
$$
C_7(m_W) =\frac{1}{2}A(x)-(\frac{f}{f'})^2[B(y)-\frac{1}{6}A(y)],
\eqno{(13)}
$$
$$
C_8(m_W)=\frac{1}{2}D(x)+(\frac{f}{f'})^2[\frac{1}{6}D(y)-E(y)],
\eqno{(14)}
$$
where functions $A$, $B$, $D$ and $E$ are defined in the Ref.[1], 
$x=(\frac{m_t}{m_W})^2$, $y=(\frac{m_t}{m_{\pi_p}})^2$.

The running of the coefficients of operators from $\mu=m_W$ to $\mu= m_b$ 
was well described in Ref.[16]. After renormalization group running we 
have the QCD corrected coefficients of operators at $\mu= m_b$ scale.
$$
C_7^{eff}(m_b) =\varrho^{-\frac{16}{23}}C_7(m_W)+\frac{8}{3}
(\varrho^{-\frac{14}{23}}-\varrho^{-\frac{16}{23}})C_8(m_W)+C_2(m_W)\sum
\limits_{i=1}\limits^{8}h_i
\varrho^{-a_i},
\eqno{(15)}
$$
with 
$$
\varrho = \frac{\alpha_s(m_b)}{\alpha_s(m_W)},
\eqno{(16)}
$$
$$
h_i=(\frac{626126}{272277}, -\frac{56281}{51730}, -\frac{3}{7}, -\frac{1}{14}
, -0.6494, -0.0380, -0.0186, -0.0057),
\eqno{(17)}
$$
$$
a_i=(\frac{14}{23}, \frac{16}{23}, \frac{6}{23}, -\frac{12}{23}, 0.4086, 
-0.4230, -0.8994, 0.1456).
\eqno{(18)}
$$

Now we write down the amplitude of Fig.1 as     
$$
\begin{array}{ll}
M_a=&\int^1_0 dx_1 dy_1 \Phi_{D_s^*}(y_1)\Phi_{B_c}(x_1)\frac{-iG_F}
{\sqrt{2}}V_{tb}
V_{ ts}^*C_7^{eff}(m_b)m_be\frac{\alpha_s(m_b)}{2\pi}C_F \\
&\{T_r[(\not q-m_{D_s}^{*})\not\epsilon\sigma_{\mu\nu}(1+\gamma_5)
k^{\nu}\eta^{\mu}(\not p-y_1\not q+m_b)\gamma_{\alpha}(\not p+m_{B_c})
\gamma_5\gamma^{\alpha}]\frac{1}{D_1D_3} \\
&+Tr[(\not q-m_{D_s^{*}})\not\epsilon\gamma_{\alpha}(\not q-x_1\not p)
\sigma_{\mu\nu}(1+\gamma_5)k^{\nu}\eta^{\mu}(\not p+m_{B_c})
\gamma_5\gamma^{\alpha}]\frac{1}{D_2D_3}\},    
\end{array}
\eqno{(19)} 
$$
where $\eta$ is the polarization vector of photon, $x_1$, $y_1$ are the 
momentum fractions shared by charms in $B_c$ and $D_s^*$ respectively, and 
the color factor $C_F$ = $\frac{4}{3}$. The factors $D_1$, $D_2$ and $D_3$ 
in eq.(19) are the forms of
$$
\begin{array}{l}
D_1=(1-y_1)(m_{ B_c}^{ 2}-m_{D_s^*}^2y_1)-m_{ b}^{ 2},{\hskip 8.5cm} (20)\\
D_2=(1-x_1)(m_{ D_s^*}^2-m_{B_c}^2 x_1),{\hskip 9.5cm}(21) \\     
D_3=(x_1-y_1)(x_1m_{B_c}^{ 2}-y_1m_{D_s^*}^{ 2}).{\hskip 9cm}(22)
\end{array}
$$

Now the amplitude $M_a$ can be written as
$$
M_a=i\varepsilon_{\mu\nu\alpha\beta}\eta^{\mu}k^{\nu}\epsilon^{\alpha}
p^{\beta}f_{ 1}^{ peng} +\eta^{\mu}[\epsilon_{\mu}(m_{B_c}^2-m_{D_s}^2)
-(p+q)_{\mu}(\epsilon\cdot k)]f_2^{ peng},     
\eqno{(23)}
$$
with form factors
$$
\begin{array}{ll}
f_1^{peng}&=2f_2^{peng}=C\int_{0}^{1}dx_1dy_1\delta(x_1-\epsilon_{B_c})
\delta(y_1-\epsilon_{D_s^*}) \\
&\{[m_{B_c}(1-y_1)(m_{B_c}-2m_{D_s^*}) 
-m_b(2m_{B_c}-m_{D_s^*})] {\frac{1}{D_1D_3}} - m_{B_c}m_{D_s^*}(1-x_1) 
{\frac{1}{D_2D_3}}\},
\end{array}   
\eqno{(24)}
$$
where
$$
C=\frac{em_bf_{B_c}f_{D_s^*}C_7^{eff}(m_b)C_F\alpha_s(m_b)G_FV_{tb}
V_{ts}^*}{12\pi\sqrt{2}}.            
\eqno{(25)}
$$    

\begin{flushleft}
\subsection*{II ( ii ).  The weak annihilation contribution}
\end{flushleft}                                             
         
As mentioned in Sec.I, $B_c$ meson is also the unique probe of 
the weak annihilation mechanism. The leading log QCD-corrected effective weak 
Hamiltonian of $W$ annihilation is
$$
H^{(W)}_{eff}= \frac{G_F}{2\sqrt{2}}V_{cb}V_{cs}^*( c_+O_+ + c_-O_- ) + H.c,  
\eqno{(26)}
$$
with $O_{\pm}=(\overline sb)(\overline cc)\pm(\overline sc)(\overline cb)$, 
where $(\overline q_1q_2)
\equiv \overline q_1\gamma_{\mu}(1-\gamma_5)q_2$, $c_{\pm}$ are Wilson 
coefficient functions.
 
Using the method developed by H. Y. Cheng $et \ \ al.$ [17], we can get the 
amplitude of $W$ annihilation diagrams ( see Fig.2 ):
$$
M_{ b}^{(W)}=i\varepsilon_{\mu\nu\alpha\beta}\eta^{\mu}k^{\nu}\epsilon^
{\alpha}
p^{\beta}f_{1(W)}^{anni}+\eta^{\mu}[\epsilon_{\mu}(m_{B_c}^2-m_{D_s}^2)-
(p+q)_{\mu}(\epsilon\cdot k)]f_{2(W)}^{anni},               
\eqno{(27)}
$$
with  
$$      
f_{1(W)}^{anni}=2\zeta[(\frac{e_s}{m_s}+\frac{e_c}{m_c})
\frac{m_{D_s^*}}{m_{B_c}}+(\frac{e_c}{m_c}+\frac{e_b}{m_b})]
\frac{m_{D_s^*}m_{B_c}}{m_{B_c}^2-m_{D_s^*}^2},            
\eqno{(28)}
$$
$$
f_{2(W)}^{anni}=-\zeta[(\frac{e_s}{m_s}-\frac{e_c}{m_c})\frac{m_{D_s^*}}
{m_{B_c}}
+(\frac{e_c}{m_c}-\frac{e_b}{m_b})]\frac{m_{D_s^*}m_{B_c}}
{m_{B_c}^2-m_{D_s^*}^2},              
\eqno{(29)}
$$       
where  
$$     
\zeta=ea_2\frac{G_F}{\sqrt{2}}V_{cb}V_{cs}^*f_{B_c}f_{D_s^*}, 
\eqno{(30)}
$$
and $a_2=\frac{1}{2}(c_--c_+)$ is a calculable coefficient in the nonleptonic 
$B$ decays.           

Using the Feynman rules given in eq. ( 1 ) and eq. ( 2 ), the leading log 
QCD-corrected effective weak Hamiltonian 
of $\pi_p^{\pm}$ annihilation is given as
$$
H^{(\pi_p)}_{eff}=-V_{cb}V_{cs}^*(\frac{f}{vf'})^2 \frac{1}{2m^2_{\pi_P}}(
c_+ O_+ + c_-O_-) + H.c.
\eqno{(31)}
$$  

Using the same method as the above, we can write down 
the amplitude of $\pi_p^{\pm}$ annihilation diagrams ( see Fig.2 ) as 
$$
M_b^{(\pi_p)}=i\varepsilon_{\mu\nu\alpha\beta}\eta^{\mu}k^{\nu}\epsilon^
{\alpha}p^{\beta}f_{1(\pi_p)}^{anni} + \eta^{\mu}[\epsilon_{\mu}(m_{B_c}^2
-m_{D_s^*}^2)-(p+q)_{\mu}(\epsilon \cdot k)]f_{ 2(\pi_p)}^{ anni},
\eqno{(32)}
$$
with      
$$
f_{1(\pi_p)}^{anni}=2\zeta^{'}[(\frac{e_s}{m_s}+\frac{e_c}{m_c})
\frac{m_s-m_c}{m_{B_c}}+(\frac{e_b}{m_b}+\frac{e_c}{m_c})
\frac{m_b-m_c}{m_{B_c}}]\frac{m_{B_c}m_{D_s^*}}{m_{B_c}^2- m_{D_s^*}^2},      
\eqno{(33)}
$$
$$  
f_{2(\pi_p)}^{ anni}=\zeta^{'}[(\frac{e_s}{m_s}+\frac{e_c}{m_c})
\frac{m_{D_s^*}}{m_{B_c}}+(\frac{e_b}{m_b}+\frac{e_c}{m_c})
]\frac{m_{D_s^{*}}m_{B_c}}{m_{ B_c}^{ 2}
-m_{ D_s^{*}}^{ 2}},        
\eqno{(34)}
$$
where                          
$$
\zeta^{'}=ea_2V_{cb}V_{cs}^*(\frac{f}{vf'})^2\frac{1}{4m_{\pi_p}^2}f_{B_c}
f_{D_s^*}(m_{B_c}^2+m_{D_s^*}^2).
\eqno{(35)}
$$

The total annihilation amplitude ( Fig.2 ) is the form of  
$$
\begin{array}{lll}
M_b&=&M_b^{(W)}+M_b^{(\pi_p)} \\
   &=&i\varepsilon_{\mu\nu\alpha\beta}\eta^{\mu}k^{\nu}\epsilon^{\alpha}
p^{\beta}f_1^{anni}+\eta^{\mu}[\epsilon_{\mu}(m_{B_c}^2-m_{D_s}^2)
-(p+q)_{\mu}(\epsilon \cdot k)]f_2^{anni},  
\end{array}
\eqno{(36)}
$$
where
$$
f_1^{anni} = f_{1(W)}^{anni} +f_{1(\pi_p)}^{anni},
\eqno{(37)}
$$
$$
f_2^{anni} = f_{2(W)}^{anni} +f_{2(\pi_p)}^{anni}.
\eqno{(38)}
$$ 

Finally, we estimate another possible long-distance effect, namely the vector
-meson-dominance ( VMD ) contribution which was advocated by Golowich and 
Pakvasa [18]. VMD implies that a possible contribution to $B_c\rightarrow 
D_s^*\gamma$ comes from the $B_c\rightarrow D_s^*J/\psi(\psi')$ followed by 
$J/\psi(\psi')\rightarrow\gamma$ conversion. 

As discussed in Refs.[19, 20], using factorization approach, 
the VMD amplitude ( Fig.3 ) is
$$
M^{VMD}=i\varepsilon_{\mu\nu\alpha\beta}\eta^{\mu}k^{\nu}\epsilon^
{\alpha}
p^{\beta}f_1^{VMD}+\eta^{\mu}[\epsilon_{\mu}(m_{B_c}^2-m_{D_s}^2)-
(p+q)_{\mu}(\epsilon\cdot k)]f_2^{VMD},               
\eqno{(39)}
$$
with  
$$      
\begin{array}{ll}
f_1^{VMD}&=eG_FV_{cb}V_{cs}^* \{\sqrt{2}a_2\frac{1}{m_{B_c}+m_{D_s^*}}
(\frac{f_{J/\psi}m_{J/\psi}}{g_{\gamma J/\psi}}+\frac{f_{\psi'}m_{\psi'}}
{g_{\gamma\psi'}})\\
& + a_1f_{D_s^*}m_{D_s^*}[\frac{1}{(m_{B_c}+m_{J/\psi})
g_{\gamma J/\psi}}+\frac{1}{(m_{B_c}+m_{\psi'})g_{\gamma\psi'}}]\} V^{BD}(0),
\end{array}
\eqno{(40)}
$$
$$
f_2^{VMD}=-\frac{1}{2}\frac{A_2^{BD}(0)}{V^{BD}(0)}f_1^{VMD},
\eqno{(41)}
$$       
where $a_1=\frac{1}{2}(c_++c_-)$, and $V^{BD}(0)$ and $A_2^{BD}(0)$ are 
form factors. 

\begin{flushleft}
\section*{III. Numerical results}
\end{flushleft}

We will use the following values for various quantities to carry on our 
calculations.
 
( i ). Decay constants for mesons. Here we use 
$$       
f_{D_s^*}=f_{D_s}=344 MeV [21], \ \ f_{B_c}=500 MeV [22], \ \ f_{J/\psi}=395 
MeV [20], \ \ f_{\psi'}=293 MeV [20].   
$$

( ii ). Meson mass and the constituent quark mass [23, 24]
$$
m_{B_c}=6.27 GeV, \ \ m_{D_s^*}=2.11 GeV, \ \ m_b=4.7 GeV, \ \ m_c=1.6 GeV,
$$
$$ 
m_s=0.51 GeV, \ \ m_{J/\psi}=3.079 GeV, \ \ m_{\psi'}=3.685 GeV.    
$$
We also use $m_{B_c}\approx m_b+m_c$, $m_{D_s^*}\approx m_s+m_c$ in our 
calculations. 

( iii ). $a_1$ and $a_2$ have been estimated very recently in Ref.[17] 
according to the CLEO data [25] on $B\rightarrow D^* \pi(\rho)$ and 
$B\rightarrow J/\psi K^*$. Here we take 
$$
a_1=1.01, \ \ a_2=0.21.
$$

( iv ). CKM matrix elements [24]. Here we use
$$
\vert V_{cb}\vert=0.04, \ \ \vert V_{ts}\vert=\vert V_{cb}\vert, \ \ \vert 
V_{cs}\vert=0.9745, \ \ \vert V_{tb}\vert=0.9991.  
$$

( v ). The QCD coupling constant $\alpha_s(\mu)$ at any renormalization 
scale, can be calculated from $\alpha_s(m_Z)=0.117$ via
$$
\alpha_s(\mu)=\frac{\alpha_s(m_Z)}{1-(11-\frac{2}{3}n_f)\frac{\alpha_s(m_Z)}
{2\pi}\ln(\frac{m_Z}{\mu})}.    
$$
We obtain 
$$
\alpha_s (m_b)=0.203,  \ \ \    \alpha_s(m_W)=0.119.                  
$$

( vi ). The Ref.[10] gives a constraint on $m_{\pi_p}$ in the 
allowed parameter space of the model: $\frac{1}{2}m_Z<m_{\pi_p}<1 TeV$. 
Here we take                         
$$
m_{\pi_p}=( 50\sim 1000 ) GeV.                        
$$

( vii ). With the constraint of $f^2+f^{'2}=v^2$ and the 
chiral perturbation theory in Ref.[10], we can get $0.115\leq \frac{f}{f'}
\leq 1.74$. Here we take 
$$	
\frac{f}{f'}=0.115
$$
in our calculations.

( viii ). The form factors $V(0)$ and $A_2(0)$ appearing in the 
two-body decays of $B$. From Ref.[26], we take
$$     
V^{BD}(0)=0.30, \ \ A_2^{BD}(0)=0.20.
$$

We present the form factors $f_i$ ( $f_1^{peng}$, $f_2^{peng}$, $f_1^{anni}$, 
$f_2^{anni}$ ) in the SM and TCMLSM in 
Table 1, so do the decay widths in Table 2 using the amplitude formula
$$
\Gamma(B_c\rightarrow D_s^*\gamma) = \frac{(m_{B_c}^2-m_{D_s^*}^2)^3}
{32\pi m_{B_c}^3}(f_1^2 +4f_2^2).               
$$

The calculated results indicate that the VMD effects are large near the pole 
and which can not be neglected. The calculated results are
$$
f_1^{VMD}=6.73\times 10^{-10}, \ \ f_2^{VMD}=-2.24\times 10^{-10},
$$
$$
\Gamma^{VMD}=1.12\times 10^{-18} GeV.
$$  

The lifetime of $B_c$ is given in Ref.[27]. In this paper we use
$$
\tau_{B_c}=( 0.4 ps\sim 1.35 ps )
$$
to estimate the branching ratio BR ( $B_c\rightarrow D_s^*\gamma$ ) which
is a function of $\tau_{B_c}$. The results are given in Table 3. 

\begin{flushleft}
\section*{IV. Discussion }
\end{flushleft}

Applying PQCD, we have studied two mechanisms which contribute to the 
process $B_c\rightarrow D_s^*\gamma$. For the short-distance one ( Fig.1 ) 
induced by electromagnetic penguin diagrams,  
the momentum square of the hard scattering exchanged by gluon is about 
$3.6 GeV^2$ which is large enough for PQCD analyzing. The hard scattering 
process can not
be included conveniently in the soft hadronic process described by the 
wave-function of the final bound state, which is one important reason that we 
can not apply the commonly used models with spectator [28] to the two body 
$B_c$ decays. There is no phase-space for the propagators appearing in 
Fig.1 to go on-shell, consequently, unlike the situation in the Ref.[6], the 
imaginary part of $M_a$ is absent. Another competitive mechanism is the weak 
annihilation. In the SM, we
 find this mechanism is as important as the former one ( they can 
contribute with the same order of magnitude ). This situation 
is distinct from that of the 
radiative weak $B^{\pm}$ decays which is overwhelmingly dominated by 
electromagnetic penguin. 
This is due to two reasons: one is that the compact 
size of $B_c$ meson enhances the importance of annihilation decays; the 
other comes from the Cabibbo allowance: in 
$ B_c \rightarrow D_s^* \gamma$, the CKM amplitude of weak 
annihilation is $\vert V_{cb}V_{cs}^*\vert $, but in $ B^{\pm} 
\rightarrow K^{\pm}\gamma$, the CKM part is $ \vert V_{ub}V_{us}^* \vert $ 
which is much smaller than $\vert V_{cb}V_{cs}^*\vert$.  

Particularly, the VMD contribution is found not small. This situation is 
quite different from the cases $B\rightarrow J/\psi K(K^*)$ and $B\rightarrow 
J/\psi\rho$ [19, 20]. The reason comes from that although the coupling of 
$J/\psi(\psi') - \gamma$ is small ( $e/g_{\gamma J/\psi(\psi')}\approx 
0.025(0.016)$ ), the $J/\psi(\psi')$ resonance effect can be very large.
 
In addition, we find that the modification of $B_c\rightarrow D_s^* \gamma $ 
from $\pi _p$ in the TCMLSM is small for the allowed range of 
mass of $\pi_p$ ( with $\frac{f}{f'}$ 
fixed ). This situation is quite different from that of Ref.[9], in which 
the size of contribution to the rare decay of $b\rightarrow s\gamma$ from 
the PGBs strongly depends on the values of the masses of the charged PGBs.   
The difference is mainly due to the small value of $\frac{f}{f'}$ 
which leads to the small modification from $\pi_p$ in the TCMLSM. However, 
in the OGTM, such suppression factor $\frac{f}{f'}$ does not exist. In our 
calculations, we take $\frac{f}{f'}$=0.115 as the input parameter. When 
$\frac{f}{f'}$ is taken properly larger ( without exceeding the constraint: 
$0.115\leq \frac{f}{f'} \leq 1.74$ ), the calculated results remain unchanged 
basically. In view of the above 
situation, it seems to indicate that the window of process $B_c\rightarrow 
D_s^*\gamma$ is close for the TCMLSM. But in our 
calculations, besides the peaking approximation of the meson wave functions, 
the theoretical uncertainties are neglected, such as that of $\alpha ( m_Z )
$, next-to-leading log QCD contribution [29], QCD correction from $m_t$ to 
$m_W$ [30], etc. When the more reliable estimation 
is available within the next few years, one can, in principle, make 
the final decision whether the window for TCMLSM is open or close.

\vspace{1cm}
\noindent {\bf ACKNOWLEDGMENT}

This work was supported in part by the National Natural Science 
Foundation of China, and by the funds from 
Henan Science and Technology  Committee.  

\vspace{1cm}
%\newpage
\begin {center}
{\bf Reference}
\end {center}
\begin{enumerate}
\item  
B. Grinstein $et \ \ al.$, Nucl. Phys. B 339 ( 1990 ) 269.
\item 
 D. S. Du and Z. Wang, Phys. Rev. D 39 ( 1989 ) 1342; K. Cheng, T. C. Yuan,  
Phys. Lett. B 325 ( 1994 ) 481, Phys. Rev. D 48 ( 1994 ) 5049;  G.R. Lu,  
Y.D. Yang and H.B. Li, Phys. Lett. B341 (1995)391, Phys. Rev. D51 (1995)2201.
\item 
J. Tang, J. H. Liu and K. T. Chao, Phys. Rev. D51 (1995)3501; K. C. 
Bowler $ et \ \ al.$, Phys. Rev. Lett. 72 ( 1994 ) 1398.
\item 
H. Leutwyler and M. Roos, Z. Phys. C 25 ( 1984 ) 91.
\item 
M. Neubert, Phys. Rep. 245 ( 1994 ) 1398.
\item 
 A. Szczepaiak $et \ \ al.$, Phys. Lett. B 243 ( 1990 ) 287; C. E. Carlson 
and 
J. Milana, Phys. Lett. B 301 ( 1993 ) 237, Phys. Rev. D 49 ( 1994 ) 5908, 
$ibid$ 51 ( 95 ) 450; H-n. Li and 
H. L. Yu, Phys. Rev. Lett. 74 ( 1995 ) 4388.
\item 
D. S. Du, $et \ \ al.$, BIHEP-TH-95-38 ( hep-ph/9603291 )
\item
S. J. Brodsky and G. P. Lepage, Phys. Rev. D 22 ( 1980 ) 2157.
\item  
Cai-Dian L\"u and Zhenjun Xiao, Phys. Rev. D 53 ( 1996 ) 2529.
\item 
C. D. Carone and H. Georgi, Phys. Rev. D 49 ( 1994 ) 1427.   
\item
 C. D. Carone and E. H. Simmons, Nucl. Phys. B 397 ( 1993 ) 591; 
E. H. Simmons, Nucl. Phys. B 312 ( 1989 ) 253; S. Samuel, $ibid$, 
B 347 ( 1990 ) 625; A. Kagan and S. Samuel, Phys. Lett. 
B 252 ( 1990 ) 605, B 270 ( 1991 ) 37.
\item 
S. J. Brodsky and C. R. Ji, Phys. Rev. Lett. 55 ( 1985 ) 2257.
\item 
Hsiang-nan Li, Phys. Rev. D 52, ( 1995 ) 3958.
\item
C. E. Carlson, J. Milana, Phys. Rev. D 51, ( 1995 ) 4950.
\item 
H. D. Politzer, Nucl. Phys. B 172 ( 1980 ) 349; H. Simma, Preprint, DESY 
93-083.
\item
M. Misiak, Phys. Lett. B 269 ( 1991 ) 161; K. Adel, Y. P. Yao, Mod. Phys. 
Lett. A 8 ( 1993 ) 1679; Phys. Rev. D 49 ( 1994 ) 4945; M. Ciuchini $ et
\ \ al.$, Phys. Lett. B 316 ( 1993 ) 127.
\item
H. Y. Cheng $et \ \ al.$, Phys. Rev. D 51 ( 1995 ) 1199.
\item
E. Golowich and S. Pakvasa, Phys. Lett. B 205 ( 1988 ) 393.
\item
M. Bauer, B. Stech and M. Wirbel, Z. Phys.C 34 ( 1987 ) 103; 
E. Golowich and S. Pakvasa, Phys. Rev. D 51 ( 1995 ) 1215. 
\item
H. Y. Cheng, Phys. Rev. D 51 ( 1995 ) 6228.  
\item 
A. Aoki $et \ \ al.$, Prog. Theor. Phys. 89 ( 1993 ) 137; D. Acosta 
$et \ \ al.$, CLNS 93/1238; J. Z. Bai $ et \ \ al.$,  BES Collaboration 
Phys. Rev. Lett. 74 ( 1995 ) 4599.
\item 
W. Buchm\"uiller and S-H. HTye, Phys. Rev. D 24 ( 1994 ) 132; A. Martin, 
Phys. Lett. B 93 ( 1980 ) 338; C. Quigg and J. L. Rosner, Phys. Lett. B 71
 ( 1977 ) 153; E. Eicheten $et \ \ al.$, Phys. Rev. D 17 ( 1978 ) 3090.
\item 
W. Kwong and J. L. Rosner, Phys. Rev. D 44 ( 1991 ) 212.
\item 
Particle Data Group, L. Montanet $et \ \ al.$,  Phys. Rev. D 50 
( 1994 ) 1173.
\item 
CLEO Collaboration, M. S. Alam $et \ \ al.$, Phys. Rev. D 50 ( 1994 ) 43.
\item
M. Neubert, CERN-TH/96-55 (hep-ph/9604412).
\item 
C. Quigg, FERMILAB-Conf-93/265-T; C. H. Chang and Y. Q. Chen, 
Phys. Rev. D49 ( 1994 ) 3399; P. Colangelo $et \ \ al.$, Z. Phys. C 57
 ( 1993 ) 43. 
\item 
M. Bauer $et \ \ al.$, Z. Phys. C29 ( 1985 ) 637; B. Grinstein $et \ \ al.$,
Phys. Rev. D39 ( 1987 ) 799.
\item
A. J. Buras $et \ \ al.$, Nucl. Phys. B 370 ( 1992 ) 69; Addendum, $ibid$, B 
375 ( 1992 ) 501, B 400 ( 1993 ) 37 and B 400 ( 1993 ) 75; M. Ciuchini $et \ 
\ al.$, Phys. Lett. B 301 ( 1993 ) 263, Nucl. Phys. B 415 ( 1994 ) 403.
\item
C. S. Gao, J. L. Hu, C. D. L\"u and Z. M. Qiu, 
Phys. Rev. D52 (1995)3978. 

\end{enumerate}

\newpage
\begin{table}[h]
 \caption{Form factors in the SM and TCMLSM. $f^{peng}$ and $f^{anni}$ 
represent form factors through electromagnetic penguin process and through 
weak annihilation process respectively.}
 \begin{center}
 \begin{tabular}{|c|c|c|} \hline
 $f_i$ & SM & TCMLSM  \\ \hline
 $f_1^{peng}$ & $-3.05\times 10^{-10}$ & $(-3.02\sim -3.05)\times 10^{-10}$
 \\ \hline
$f_2^{peng}$ & $-1.52\times 10^{-10}$ &$ -1.52\times 10^{-10}$ 
 \\ \hline
$f_1^{anni}$ & $7.10\times 10^{-10}$ & $ 7.10\times10^{-10}$ 
 \\ \hline
$f_2^{anni}$ & $-1.70\times 10^{-10}$ & $ -1.70\times 10^{-10}$ 
 \\ \hline    
\end{tabular}
 \end{center}
 \end {table}

\begin{table}[h]
\caption{The decay rates in the SM and TCMLSM. The $\Gamma^{peng}$, $\Gamma^
{anni}$ and $\Gamma^{total}$ represent  $\Gamma$ ( $B_c\rightarrow D_s^*
\gamma$ ) through electromagnetic penguin process, through weak annihilation 
process and penguin + annihilation respectively.}
 \begin{center}
 \begin{tabular}{|c|c|c|}  \hline
 $\Gamma(B_c \rightarrow D_s^* \gamma)$  & SM  & TCMLSM  \\ \hline
 $\Gamma^{peng}$(GeV)  &$ 3.18 \times 10^{-19}$ & $ ( 3.12\sim 3.17 )
\times 10^{-19}$  \\ \hline
 $\Gamma^{anni}$(GeV)  & $1.06\times 10^{-18}$ & $ 1.06
\times 10^{-18}$  \\ \hline
$\Gamma^{total}$(GeV)  & $4.03\times 10^{-18}$ & $ 4.03\times 
10^{-18}$  \\ \hline
 \end{tabular}
 \end{center}
 \end{table}

\begin{table}[h]
\caption{The branching ratio ( $B_c\rightarrow D_s^*\gamma$ ). The $BR^{SM}_
{total}$ and $BR^{TCMLSM}_{total}$ represent the branching ratio ( $B_c
\rightarrow D_s^*\gamma$ ) in the SM and TCMLSM respectively.}
\begin{center}
\begin{tabular}{|c|c|c|c|} \hline
$\tau_{B_c} $ & 0.4ps & 1.0ps & 1.35ps  \\ \hline
$ BR^{SM}_{total} $ & $2.44\times 10^{-6}$ & $
6.12\times 10^{-6}$ & $8.27\times 10^{-6}$ \\ \hline
$BR^{TCMLSM}_{total}$ &$2.44\times 10^{-6}$ 
&$6.12\times 10^{-6}$ &$8.27\times 10^{-6}$ \\ \hline
\end{tabular}
\end{center}
\end{table} 

\vspace{1cm}
%\newpage
\begin{center}
{\bf Figure captions}                    
\end{center}

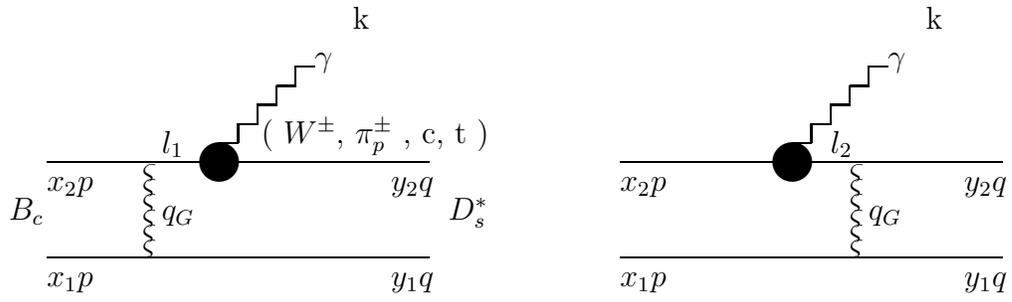
Fig.1: The Feynman diagrams which contribute to the rare radiative 
decay $B_c\rightarrow D_s^*\gamma$ through electromagnetic penguin process. 
The blob represents the electromagnetic penguin operators contributing to 
$b\rightarrow s\gamma$, $x_2 p$ and $x_1 p$ are momenta of b and c quarks in 
the $B_c$ meson respectively, $y_2 q$ and $y_1q$ are momenta of s and c 
quarks in the $D_s^*$ meson, respectively.

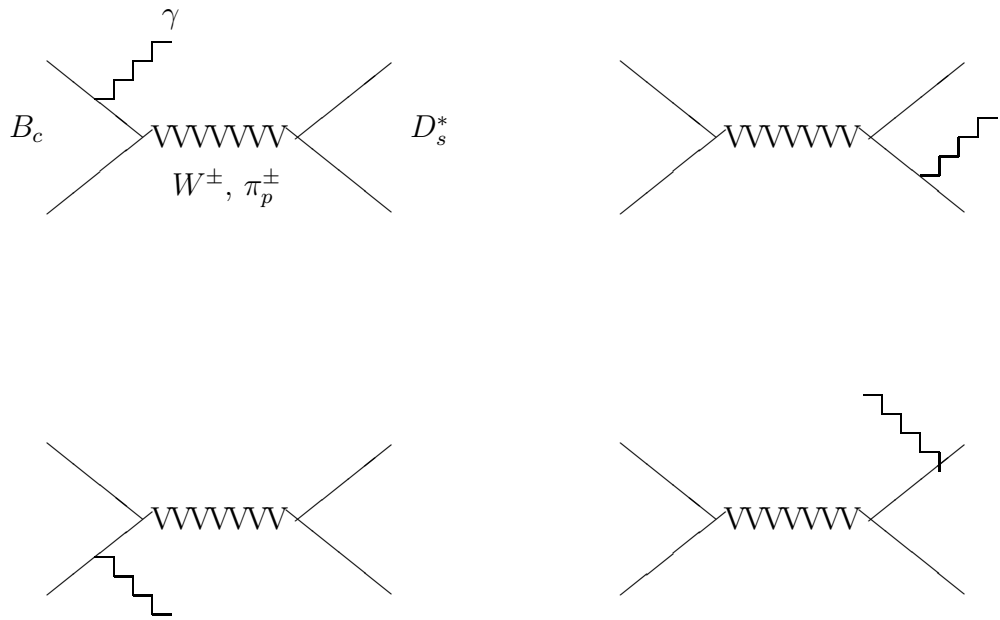
Fig.2: The Feynman diagrams which contribute to the rare radiative 
decay $B_c\rightarrow D_s^*\gamma$ through weak annihilation process. In the 
SM, there is only $W^{\pm}$ annihilation, in the TCMLSM, there are both 
$W^{\pm}$ and $\pi_p^{\pm}$ annihilations.        

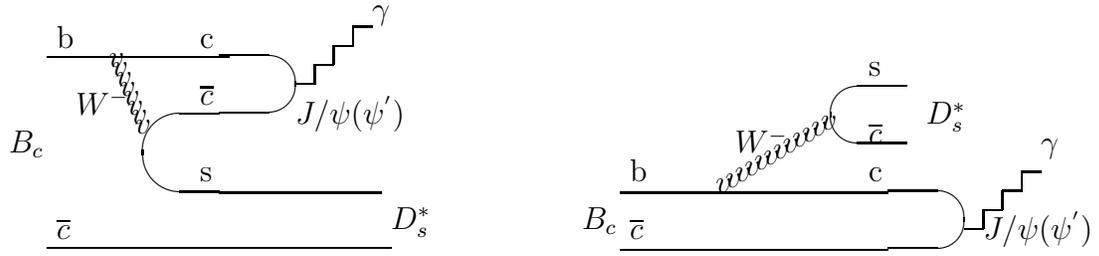
Fig.3: VMD processes which contribute to $B_c\rightarrow D_s^*\gamma$ 
with the vector-meson intermediate states $J/\psi(\psi')$.

\newpage
%%
%%
%%
%%%%%%%%%%%%%%%%%   3. the Latex files of the figures  %%%%%%%%%%%%%%
%%
%%
\begin{picture}(30,0)

\setlength{\unitlength}{0.1in}
%1
\put(5,-15){\line(1,0){20}}
\put(5,-20){\line(1,0){20}}
 \put(3,-18){$B_c$}  
  \put(26,-18){$D_s^*$}
   \multiput(10,-15.8)(0,-0.8){6}{$\varsigma$}
   \multiput(14,-15)(1,1){5}{\line(0,1){1}} 
   \multiput(13,-15)(1,1){6}{\line(1,0){1}}
    \put(14,-15){\circle*{3}}
      \put(11,-18){$q_G$}
\put(16.2,-14){( $W^{\pm}$, $\pi_p^{\pm}$ , c, t )}
\put(5,-16.5){$x_2 p$}
\put(5,-21.5){$x_1 p$}
\put(23,-16.5){$y_2 q$}
\put(23,-21.5){$y_1 q$}
\put(11,-14.5){$l_1 $}
\put(29,-25.2){Fig.1}
\put(21,-8){k}
\put(19,-10){$\gamma$} 
%2 
\put(35,-15){\line(1,0){20}}
\put(35,-20){\line(1,0){20}}
   \multiput(47,-15.8)(0,-0.8){6}{$\varsigma$}
   \multiput(44,-15)(1,1){5}{\line(0,1){1}} 
   \multiput(43,-15)(1,1){6}{\line(1,0){1}}
\put(48,-18){$q_G$}
\put(35,-16.5){$x_2 p$}
\put(35,-21.5){$x_1 p$}
\put(53,-16.5){$y_2 q$}
\put(53,-21.5){$y_1 q$}
\put(46,-14.5){$l_2 $}
\put(44,-15){\circle*{3}} 
\put(51,-8){k}
\put(49,-10){$\gamma$}
%3
\put(5,-35){\line(5,-4){5}}
\put(5,-43){\line(5,4){5.5}}
\multiput(10.4,-39.5)(1,0){7}{V}
\put(17.5,-38.5){\line(5,-4){5.5}}
\put(18,-39.1){\line(5,4){5}}
   \multiput(51.7,-41)(1,1){3}{\line(0,1){1}} 
   \multiput(50.7,-41)(1,1){4}{\line(1,0){1}}
 \put(3,-39){$B_c$}  
  \put(24,-39){$D_s^*$}
\put(11,-42){ $W^{\pm}$, $\pi_p^{\pm}$ }
\put(11,-33){$\gamma$}
%4 
\put(35,-35){\line(5,-4){5}}
\put(35,-63){\line(5,4){5.5}}
\multiput(40.4,-59.5)(1,0){7}{V}
\put(47.5,-58.5){\line(5,-4){5.5}}
\put(48,-59.1){\line(5,4){5}}
\multiput(8.5,-61)(1,-1){3}{\line(0,-1){1}} 
\multiput(7.5,-61)(1,-1){4}{\line(1,0){1}}
%5
\put(5,-55){\line(5,-4){5}}
\put(5,-63){\line(5,4){5.5}}
\multiput(10.4,-59.5)(1,0){7}{V}
\put(17.5,-58.5){\line(5,-4){5.5}}
\put(18,-59.1){\line(5,4){5}}
   \multiput(51.7,-55.5)(-1,1){4}{\line(0,-1){1}} 
   \multiput(50.7,-55.5)(-1,1){4}{\line(1,0){1}}
%6 
\put(35,-55){\line(5,-4){5}}
\put(35,-43){\line(5,4){5.5}}
\multiput(40.4,-39.5)(1,0){7}{V}
\put(47.5,-38.5){\line(5,-4){5.5}}
\put(48,-39.1){\line(5,4){5}}
\multiput(8.5,-37)(1,1){3}{\line(0,1){1}} 
\multiput(7.5,-37)(1,1){4}{\line(1,0){1}}
\put(29,-70){Fig.2} 

\end{picture}

\newpage

\begin{picture}(30,0)

\setlength{\unitlength}{0.1in}

%7
\put(5,-15){\line(1,0){9.5}}
\put(14,-16.4){\oval(8,3)[r]}
\multiput(9.6,-19)(-0.2,0.5){8}{$v$}
\put(6.5,-18){$W^-$}
\put(14,-20){\oval(8,4.1)[l]}
\put(5,-25){\line(1,0){18}}
\put(14,-22.1){\line (1,0){8.5}}
\put(18,-18.5){$J/{\psi}(\psi^{'})$}
\multiput(19,-16.4)(1,1){3}{\line(0,1){1}}
\multiput(18,-16.4)(1,1){4}{\line(1,0){1}}
\put(22,-13){$\gamma$}
\put(5.5,-14.5){b}
\put(13,-14.5){c}
\put(13,-21.5){s}
\put(13,-17.5){$\overline {c}$}
\put(5.5,-24.5){$\overline {c}$}
\put(3,-20){$B_c$}
\put(23,-24){$D_s^*$}
\put(29,-30){$Fig.3$}
%8
\multiput(40,-22.1)(0.5,0.3){12}{$v$}
\put(41,-20){$W^-$}
\put(50,-18){\oval(8,3)[l]}
\put(35,-25.1){\line(1,0){14}}
\put(35,-22.1){\line (1,0){14}}
\put(49,-23.5){\oval(8,3)[r]}
\put(54,-24.5){$J/{\psi}(\psi^{'})$}
\multiput(54,-24)(1,1){3}{\line(0,1){1}}
\multiput(53,-24)(1,1){4}{\line(1,0){1}}
\put(57,-20){$\gamma$}
\put(35.5,-21.5){b}
\put(48,-21.5){c}
\put(48,-16){s}
\put(48,-19.5){$\overline {c}$}
\put(35.5,-24.5){$\overline {c}$}
\put(33,-24){$B_c$}
\put(51,-18.5){$D_s^*$}

\end{picture}

\end{document}